# Orbital Angular Momentum Antennas: Understanding Actual Possibilities Through the Aperture Antennas Theory

Andrea Francesco Morabito[1], Loreto Di Donato[2], and Tommaso Isernia[1]

[1] DIIES Department of Information, Infrastructures and Sustainable Energy, University of Reggio Calabria, Via Graziella, Loc. Feo di Vito, I-89124 Reggio Calabria, Italy (email: andrea.morabito@unirc.it; tommaso.isernia@unirc.it)

[2] DIEEI Department of Electrical, Electronics and Informatics Engineering at University of Catania, Viale A. Doria 6, I-95124 Catania, Italy (email: loreto.didonato@dieei.unict.it)

*Abstract*—**We propose a simple method based on the Aperture Antennas Theory to understand in a simple fashion limitations of OAM antennas in far-field links. Additional insight is also given by analyzing the properties of the operators relating source and far-field distributions for a given order of the vortex. The outcomes fully agree with the results recently achieved by Edfors, Craeye, and co-authors, and emphasize some additional draw-back. The 'degrees of freedom' of the fields associated to the different orders of the vortices are also discussed.**

## I. A PREMISE

Amongst the novel techniques developed for utilizing the radio spectrum with maximum efficiency, great attention has been recently devoted to Orbital Angular Momentum (OAM) antennas [1]-[20]. Such systems have been in some cases proposed as means to improve almost "indefinitely" the channel capacity in a link amongst two antennas [2].

Roughly speaking, the idea is to take profit from the fact that an antenna could simultaneously generate different fields each one associated to a different amount of 'orbital momentum', i.e., to a different angular variation of the phase around the target direction (such as $e^{j\phi}$, $e^{2j\phi}$, and so on, $\phi$ denoting the azimuth angle in the observation domain). Then, by associating a different information to each of these patterns, one could realize an "OAM multiplexing" [17] and eventually enlarge "at will" the channel capacity [2]. Saying it in other words, it is like using different modes of a channel, where the modes are associated to a free space link rather than to a guiding structure.

A large interest has been devoted to such a topic, including spectacular public demonstrations for mass media (e.g., Piazza San Marco, Venice, Italy, June 24th 2011 [2]) as well as contributions on top scientific journals, e.g., [3]-[6].

Since then, doubts [7] and objections [8],[10] have begun to appear, emphasizing some expected limitations.

In the attempt to contribute to such a debate, in the following we propose a very simple yet instructive point of view on the subject. In particular, we focus on the actual possibility to get such a multiplication of channels [2] in the far-field zone, which is the usual framework for antenna links. In this respect, we will focus on the case of aperture antennas. Notably, this assumption does not impair the general validity of the following discussion. In fact, any antenna can be regarded as an aperture antenna by a proper choice of the aperture plane [21]. Also, the arguments below could be rephrased in terms of multipole expansions as well as in terms of the appropriate radiation integrals. Then, additional insight is also given by an analysis of the operators relating source and field behaviors for a given order of the vortex. The outcomes confirm limitations indicated in [7],[8], and emphasize some further draw-backs of the proposed multiplexing scheme.

The paper is organized as follows. In Sections II and III, respectively, we review some properties of the Hankel transform and exploit them to prove a couple of fundamental limitations of OAM antennas. Then, in Section IV, we provide an even deeper understanding by analyzing the properties of the operators relating the far field 'vortices' to the corresponding sources. Finally, the degrees of freedom' of the ℓ-th order vortex are discussed, and examples witnessing typical behaviors and performances are given in Section V. Conclusions follow.

## II. Understanding limitations through Hankel transforms

As the far field of an aperture antenna is proportional (but for a kind of 'element factor') to the Fourier transform of the aperture field [21], a very convenient mathematical setting is the one suggested in [22],[23]. The latter is based on the exploitation of Hankel transforms, whose properties are extensively analyzed and discussed in [24],[25].

Hankel transforms have already been used in [12] in order to understand some OAM antennas' limitations. However, the following analysis is completely different and provides a number of novelties with respect to [12]. In fact, we will provide conclusions which are based either on the kernel of the involved operators (Section III) or on a Singular Value Decomposition (SVD) of these latter (Section IV).

With respect to our specific problem, such a tool can be exploited as follows.

Let $f(\rho',\phi')$ denote the component of interest of the aperture field, with $\rho'$ and $\phi'$ respectively being the radial and angular variables spanning the aperture (which is supposed to be circular and located in the $xy$ plane, see Fig. 1). By virtue of (9.14) in [25], $f(\rho',\phi')$ can be expanded in a multipole series as:

$$f(\rho',\phi') = \sum_{\lambda=-\infty}^{\infty} f_\lambda(\rho') e^{j\lambda\phi'} \qquad (1)$$

where:

$$f_\lambda(\rho') = \frac{1}{2\pi} \int_0^{2\pi} f(\rho',\phi') e^{-j\lambda\phi'} d\phi' \qquad (2)$$

If $k'$ and $\phi$ respectively denote the radial and azimuth coordinates in the spectral domain, then the Fourier transform of the source (1) is given by {see (9.15) in [25]}:

$$F(k',\phi) = \frac{1}{2\pi} \int_0^{2\pi}\int_0^\infty f(\rho',\phi') e^{-jk'\rho'\cos(\phi'-\phi)} \rho' d\rho' d\phi' \qquad (3)$$

which can in turn be expanded in a multipole series as follows:

$$F(k',\phi) = \sum_{\lambda=-\infty}^{\infty} F_\lambda(k') e^{j\lambda\phi} \qquad (4)$$

where:

$$F_\lambda(k') = \frac{1}{2\pi} \int_0^{2\pi} F(k',\phi) e^{-j\lambda\phi} d\phi \qquad (5)$$

Finally, by substituting (1) into (3) and then using (4), one achieves {see (9.17) in [25]}:

$$F_\lambda(k') = \int_0^\infty f_\lambda(\rho') J_\lambda(k'\rho') \rho' d\rho' = H_\lambda\{f_\lambda(\rho')\} \qquad (6)$$

where $J_\ell$ is the $\ell$-th order Bessel function of first kind, and expression (6) is indeed the Hankel transform [24] of order $\ell$ of the function $f_\ell(\rho')$.

As well known, the 'visible' part of the spectrum, i.e., $k' \leq \beta$ ($\beta=2\pi/\lambda$ denoting the wavenumber, with $\lambda$ the wavelength) determines the actual far-field behavior. In particular, by denoting with $\theta$ the elevation angle with respect to the boresight and adopting the usual correspondence with the spectral variables $u=\beta\sin\theta\cos\phi$ and $v=\beta\sin\theta\sin\phi$, one finds:

$$k' = \sqrt{u^2+v^2} = \beta\sin\theta \qquad (7)$$

Equations (1)-(7) imply a number of simple yet interesting consequences:

i. an angular variation of order $\ell$ of the source in terms of the $\phi'$ variable corresponds to an angular variation of order $\ell$ of the far field in terms of the $\phi$ variable. Hence, a natural diagonalization of the relationship between the aperture field and the corresponding spectrum occurs;

ii. for any fixed order $\ell$ of angular variation, the function $f_\ell(\rho')$ univocally determines the corresponding function $F_\ell(k')$ and vice versa. Both the forward and backward relations are ruled by a Hankel transform of order $\ell$;

iii. the Hankel-transform relationships (6) determines (but for a slowly-varying factor) the far-field power pattern associated to each source component $f_\ell(\rho') e^{j\ell\phi}$.

The last point above can also be considered in the reverse fashion: once a desired power pattern is specified in terms of $\theta$ and $\phi$, and some reasonable prolongation is used for the invisible part of the spectrum [26], relations above univocally define the corresponding source. In particular, if a given angular variation of the kind $e^{j\ell\phi}$ is desired in conjunction with a given elevation behavior specified by some $F_\ell(\beta\sin\theta)$ function, then the inverse Hankel transform will allow determining the corresponding source. In fact, after prolonging the $F_\ell(k')$ function in the invisible part of the spectrum, by using the inverse Hankel transform {see (9.18) in [25]}, the sought source will be given (but for the implicit angular variation) by:

$$f_\lambda(\rho') = \mathrm{H}_\lambda^{-1}\{F_\lambda(k')\} = \int_0^\infty F_\lambda(k')J_\lambda(k'\rho')k'\,dk' \qquad (8)$$

## III. A COUPLE OF STRAIGHTFORWARD CONSEQUENCES

With the aim of getting some preliminary understanding, let us look to the forward (6) and backward (8) relationships. In particular, let us have a look of what is going to happen with increasing values of $\ell$ and limited sizes of the source.

As a crucial (but often overlooked) circumstance, $\forall \ell \neq 0$ Bessel functions $J_\ell$ present a $|\ell|$-th order zero in the origin [27]. Moreover, the $J_\ell$ function is the unique term of (6) depending on $k'$. Then, whatever the source at hand, $\forall \ell \neq 0$ the corresponding spectrum (and hence the far field) will have a 'hole' (or better, a null) in the boresight direction while still being different from zero elsewhere. Notably, because of the properties of Bessel functions [see also (10) below], such a hole will have increasing size and depth with $|\ell|$.

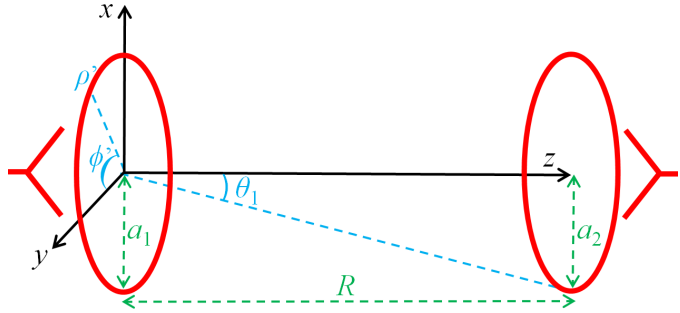

Fig. 1. Simple communication link wherein the receiving aperture antenna (on the right) is located in the broadside direction of the transmitting aperture antenna (on the left).

Under such circumstances, even assuming a receiver positioned at the broadside direction (see Fig. 1) is able to detect and understand the (weaker and weaker) signal associated to the $\ell$-th order vortex, a huge price is paid. In fact, the majority of the power radiated by the transmitting antenna is spread out of the broadside direction, with two related consequences. First, the field level is uselessly large in a number of spatial directions, potentially violating radiation limits. Second, spectral resources are wasted. In fact, a number of potential channels based on space re-use (through multibeam antennas) are occupied by the toroidal power patterns associated to the vortices.

Therefore, unless the receiving antenna is so large to intercept the power maxima, one will have a waste of both power and potential space-diversity-based channels, as well as an unjustified electromagnetic pollution.

As a further draw-back, $\forall |\ell| > 0$ the received signal will undergo a very rapid decrease with the distance. In fact, if $R$ denotes the link distance while $a_2$ and $\theta_1$ respectively denote the radius of the receiving antenna and the elevation angle corresponding to its borders, one will have (see Fig. 1):

$$\sin\theta_1 = \frac{a_2}{\sqrt{R^2 + a_2^2}} \qquad (9.\text{a})$$

Then, if $R \gg a_2$ (which is verified in the far-field region), it is:

$$\sin\theta_1 \approx (a_2/R) \ll 1 \qquad (9.\text{b})$$

as a consequence, in the overall cone $0 \leq \theta \leq \theta_1$, using (9.1.7) of [27], it is:

$$\left|J_\lambda(\rho'\beta\sin\theta)\right| \approx \frac{1}{|\lambda|!}\left(\frac{\rho'\beta\sin\theta}{2}\right)^{|\lambda|} \qquad (10)$$

$$\forall \lambda: \quad 0 < \rho'\beta\sin\theta \ll \sqrt{|\lambda| + 1}$$

so that the borders of the antenna correspond to the maximal power density which can be intercepted.

Notably, (9.b) and (10) also entail that in the overall cone $0 \leq \theta \leq \theta_1$ the usual $(1/R)^2$ power attenuation will be complemented by a $(1/R)^{2|\ell|}$ factor for the $\ell$-th order vortex field. Such a characteristic of $\ell$-th order vortices has been already recognized and emphasized (on the basis of a completely different argument) in [8] as well as (previously) in [28].

## IV. FURTHER TOOLS AND ANALYSIS

Additional insights into the analysis carried out in the previous Section can be achieved by exploiting the Singular Value Decomposition (SVD) [29] of the relevant operators (6) (as truncated over a finite circular domain). SVD has been already used in the study of OAM antennas in [7], wherein authors compared the OAM antennas' performance to that of traditional Multiple-In-Multiple-Out (MIMO) systems by applying the SVD to the channel matrix of the latter.

Herein, the use of SVD is completely different. In fact, we analyze the behavior of finite-dimensional sources by performing the SVD of the $\ell$-th order relevant operators (6). By so doing, we will be able to define which kinds of far field can be realized by exploiting given angularly-varying aperture fields, as well as how much energy is needed for generating angularly-varying far fields having a given $\ell$ value while possibly guaranteeing given performances in the broadside direction.

In order to apply the SVD to (6), it proves convenient to re-write the latter in the case of a generic source defined over a circular aperture of limited radius $a$, and to provide some renormalizations of the aperture disk and of the visible region. By so doing, one achieves (apart from inessential constants):

$$F_\lambda(k) = \int_0^1 f_\lambda(\rho) J_\lambda(\beta a k \rho) \rho d\rho \qquad (11)$$

with $\rho = \rho'/a$ and $k = k'/\beta$ (so that both the normalized variables $\rho$ and $k$ belong to the interval [0,1]).

Equation (11) leads to the following formulation in terms of the operator $A_\ell$:

$$F_\lambda = A_\lambda f_\lambda \qquad (12.a)$$

where:

$$A_\lambda : f_\lambda \in L_2(0,1) \rightarrow A_\lambda f_\lambda \in L_2(0,1) \qquad (12.b)$$

Let us now denote by $\{v_{\ell,n}, \sigma_{\ell,n}, u_{\ell,n}\}$ the SVD of $A_\ell$, i.e., the functions and scalars such that:

$$A_\lambda v_{\lambda,n} = \sigma_{\lambda,n} u_{\lambda,n}$$
$$A_\lambda^+ u_{\lambda,n} = \sigma_{\lambda,n} v_{\lambda,n} \qquad (13)$$

wherein $A_\ell^+$ is the adjoint of $A_\ell$ while $\sigma_{\ell,n}$, $v_{\ell,n}$, and $u_{\ell,n}$ respectively denote the $n$-th singular value, right-hand singular function, and left-hand singular function associated to the $\ell$-th OAM mode.

For any fixed size of the source, the SVD (13) can be computed by applying the theory and formulas given in Appendix I and Appendix II, wherein connections with the Generalized Prolate Spheroidal Wave (GPSW) functions discussed in [22],[23] are also given.

Notably, the singular functions are orthonormal in the spaces of sources (i.e., $\rho \leq 1$) and far fields (i.e., $|k| \leq 1$), respectively, so that

$$\int_0^1 v_{\lambda,n}(\rho) v_{\lambda,p}^*(\rho) \rho d\rho = \int_0^1 u_{\lambda,n}(k) u_{\lambda,p}^*(k) k dk = \delta_{n,p} \quad \forall \lambda \qquad (14)$$

wherein $\delta$ denotes the Kronecker delta function and * means complex conjugation. Hence, they can act as a basis in the respective domain. Notably, they also allow a diagonalization of the relationship between sources and corresponding far fields. In fact, one can expand the aperture field and the visible spectrum as:

$$f_\lambda(\rho) = \sum_{n=1}^{\infty} \alpha_{\lambda,n} v_{\lambda,n}(\rho) \qquad (15.a)$$

$$F_\lambda(k) = \sum_{n=1}^{\infty} b_{\lambda,n} u_{\lambda,n}(k) \qquad (15.b)$$

where a suitable truncation of expansions, which affects the rate of variation of the aperture field along the radial coordinate, will be used in actual instances. In particular (see below) truncation will be needed to regularize the inversion procedure from the desired far field to the aperture field in order to circumvent superdirectivity problems [30]-[32].

Notably, because of the theory in Appendix I, the $b_{\ell,n}$ coefficients are simply related to the $\alpha_{\ell,n}$ coefficients as follows:

$$b_{\lambda,n} = \sigma_{\lambda,n} \alpha_{\lambda,n} \qquad (15.c)$$

In fact, the singular values can be thought as scalar (gain) factors by which each source is multiplied to give the corresponding far field.

Notably, (15) provide two different (related) ways for understanding some important OAM antennas' limitations.

First, one can observe the behavior of the singular values and singular functions associated to a given source size and different vortex orders. In this respect, it is worth noticing that thanks to the properties of the relevant operators, the left singular functions (corresponding to the sources) and the right singular functions (corresponding to the radiated far fields) have a similar behavior.

Second, one can solve an ad-hoc synthesis problem with reference to the simple communication link depicted in Fig. 1, wherein a receiving aperture antenna (see Section II for the definition of the parameters $a_2$, $R$, and $\theta_1$) is located in front of a transmitting OAM antenna having a radius equal to $a_1$. In particular, we are interested in understanding how much power is required on the aperture for getting an equal power at the receiving points (for identical collecting areas).

Notably, although contributions relying on Hankel transform already exist (see for instance [12]), it is indeed the first time that diagonalization (15) is exploited in order to investigate the OAM antennas' actual potentialities.

## V. ANALYSIS AND SYNTHESIS OF OAM ANTENNAS

By exploiting the tools above, we discuss herein the typical outcomes experienced in the analysis and synthesis of OAM antennas. In particular, in Subsection V.A we show and comment the singular values and singular functions associated to OAM sources generating different OAM modes, while in Subsection V.B we present the results achieved by solving the synthesis problem mentioned at the end of the previous Section for different values of $\ell$, $a_1=a_2$, and $R$. In all cases, at least 60 points per wavelength have been used in both the spatial and spectral domains in order to discretize the radiation operators and computing the corresponding singular values and singular functions.

*V.A The forward problem: singular values and singular functions of the different radiation operators $A_\ell$*

We computed the singular values and the singular functions of the operator $A_\ell$ for different values of $\ell$ and for different sizes of the aperture. As a typical behavior, the singular values corresponding to the cases $a_1=4\lambda$ and $a_1=10\lambda$ are shown for $\ell=0,1,3,5,7$ in Fig. 2 [subplots (a) and (b), respectively].

As it can be seen, in agreement with the theory in [33,34], for any fixed value of $\ell$ the singular values associated to OAM exhibit a step-like behavior, with an exponential decay after a given value of the index $n$ (let us say $N_\ell$). Then, if $\ell$ and $n$ are such that $n>N_\ell$, realization of the field $u_{\ell,n}$ implies [because of (15.c)] a very strong increase of the coefficient $\alpha_{\ell,n}$ (and hence of the energy of the corresponding source). Moreover, whenever $n>N_\ell$ the corresponding source is extremely fast oscillating, thus leading to a high-energy, fast oscillating (and hence difficult to be realized) source behavior. As fast oscillations imply a large content in the invisible part of the spectrum, and hence a very high $Q$ factor [31], such a source would also have a very narrow band.

It can also be noted from Fig. 2(a) that $N_\ell$ is smaller and smaller for increasing values of $|\ell|$. Such a circumstance implies that the larger $|\ell|$, the lower the number of linearly independent patterns which can be realized by finite-energy sources. This is not surprising, as for increasing $|\ell|$ values the fields have to accommodate wider and wider 'holes' in front of the transmitting antennas [see for example Fig. 2(d)], while necessarily being *bandlimited* elsewhere [35].

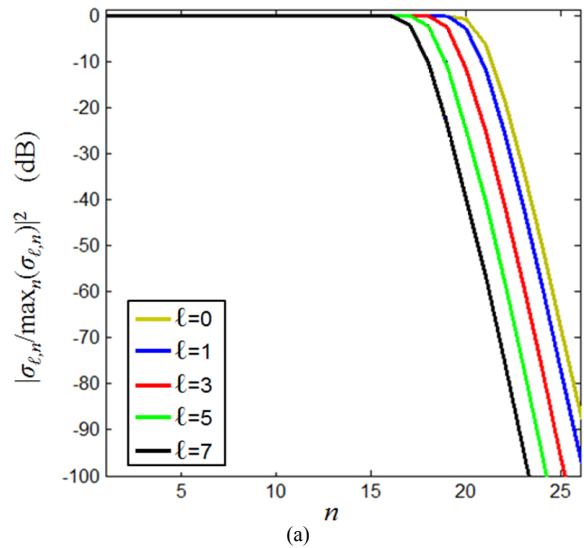

(a)

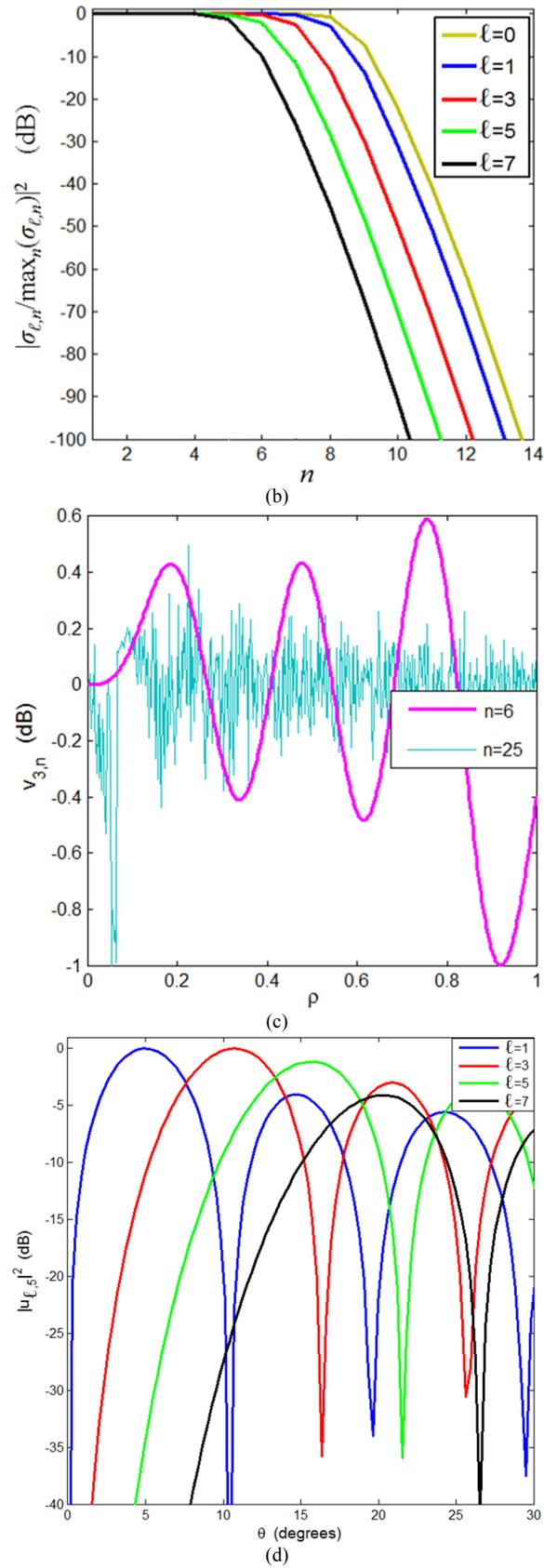

Fig. 2. Singular values associated the radiation operator for different values of $\ell$ and $n$, with $a_1=10\lambda$ [subplot (a)] and $a_1=4\lambda$ [subplot (b)]. Comparison for different $n$ values of the right-hand singular functions corresponding to $\ell=3$ and $a_1=4\lambda$ [subplot (c)]. Comparison for different $\ell$ values of the left-hand singular functions corresponding to $n=5$ and $a_1=4\lambda$ [subplot (d)].

As a final comment, let us note that all the analysis is fully coherent with the general theory of the 'degrees of freedom' of the fields radiated by finite-dimensional sources given in [35]. In fact, our numerical experiments (performed for very many possible sizes of the source) and theoretical arguments (see below) suggest that the values of $N_\ell$ obey the following rule:

$$N_\lambda = \frac{1}{\pi}\left(\beta a - |\lambda|\right) = \frac{2a}{\lambda} - \frac{|\lambda|}{\pi} = N_0 - \frac{|\lambda|}{\pi} \qquad (16)$$

In fact, the first addendum is equal to $a/(\lambda/2)$, which is known to be (see [36]) the number of degrees of freedom associated to a circularly-symmetric source of radius $a$. Moreover, as already discussed, the circumstance that the field must have an $\ell$-th order zero in the origin suggests that $N_\ell$ must decrease with $|\ell|$. Both results in Fig. 2 (and many others) as well as an a-posteriori check (see below) suggest that $1/\pi$ is the correct coefficient for such a decrease.

The rule (16) has two crucial consequences.

First, the maximum-order vortex which can be excited by a non-superdirective source, i.e., *the maximum feasible value of $\ell$*, is:

$$\lambda_{MAX} = 2\pi a/\lambda \qquad (17)$$

which is achieved by enforcing $N_{\lambda_{MAX}} = 0$.

Second, the source's overall *number of degrees of freedom*, which is equal to the sum of all feasible $N_\ell$ values, results:

$$N_{TOT} = \sum_{|\lambda|\le\lambda_{MAX}} N_\lambda =$$
$$= N_0 + 2\sum_{\lambda=1}^{\lambda_{MAX}} N_\lambda = N_0 + 2\sum_{\lambda=1}^{2\pi a/\lambda}\left(\frac{2a}{\lambda} - \frac{\lambda}{\pi}\right) = \qquad (18)$$
$$= \frac{2a}{\lambda} + 2\frac{2a}{\lambda}\frac{2\pi a}{\lambda} - \frac{2}{\pi}\sum_{\lambda=1}^{2\pi a/\lambda}\lambda = \frac{4\pi a^2}{\lambda^2}$$

This number is identical to the one given by formula (26) of [36], i.e.:

$$N_{TOT} = \frac{Area\ of\ the\ source}{(\lambda/2)^2} = 4\pi\left(\frac{a}{\lambda}\right)^2 \qquad (19)$$

which also provides an a-posteriori check of (16).

Of course, all these considerations go against the pretension [2],[9],[11] of arbitrarily enlarging the channel capacity. In fact, fields exhibiting azimuthal variations of arbitrarily-large order are de facto unfeasible when using finite-dimensional sources.

A further discussion of actual possibilities is given with reference to a specific synthesis problem in the following Subsection.

*V.B Synthesis of the aperture field required to generate a vortex field having a given intensity at the receiver*

In order to understand actual possibilities of multiplexing through vortices, we have solved, for different values of $\ell\neq0$, the following Convex Programming (CP) problem in the unknowns $\alpha_{\ell,1}$, ..., $\alpha_{\ell,N}$:

$$Minimize: \quad P_{AP}\left(\alpha_{\lambda,1},...,\alpha_{\lambda,N}\right) = \int_0^1 \left|f_\lambda(\rho)\right|^2 \rho d\rho \qquad (20.a)$$

$$Subject\ to: \quad \mathrm{Re}\left[F_\lambda(\theta_1)\right] \ge 1 \qquad (20.b)$$

$$\mathrm{Im}\left[F_\lambda(\theta_1)\right] = 0 \qquad (20.c)$$

where the functional in (20.a), which is a real and positive quadratic form of the unknowns, is equal to the aperture power associated to transmitting antenna (but for a $2\pi$ factor), while the linear constraints (20.b) and (20.c) guarantee a spectrum amplitude of at least 0 dB for $\theta=\theta_1$. Therefore, the aim of the synthesis problem (20) is that of identifying the minimum energy required in order to generate a far field having a given $\ell$-th order and being able, at the same time, to guarantee a given signal intensity at the edge of the receiving antenna. As the fields received (at the borders) can be properly recombined in order to maximize the received signal, solution of the proposed synthesis problem also will allow comparing the efforts required to have an identical received power (through equal collecting areas)[1].

---

[1] Because of the presence of larger and larger holes in the spectrum for $0\le\theta\le\theta_1$ and $|\ell|\neq0$, consideration of the power collected over the whole receiver's area (rather than $|F_1(\theta_1)|$) imply a worsening of the actual high-order vortices' performances. Therefore, figures 3 and 4 provide an 'optimistic' estimation of OAM antennas' capabilities.

In the different trials, in order to avoid superdirective sources (see above) we used a truncation of (15.a),(15.b) by adopting an index slightly larger than $N_\ell$. Then, we solved problem (20) for $\ell=1$, $\ell=2$, and $\ell=3$ and for different values of $a_1=a_2$. Concerning the value of $R$, we limited the analysis to distances not much larger than the Fraunhofer value $R_f=8a_1^2/\lambda$. This choice has been induced by the circumstance that the OAM antennas' performance in terms of ratio between received and transmitted powers decays as $R^{-2(|\ell|+1)}$ {see (20) in [16], and also [7],[8],[10],[12]}. For each combination of $\ell$, $a_1=a_2$, $N_\ell$, and $R$, in order to evaluate the energy necessary to generate a particular vortex order, the achieved $P_{AP}$ value has been compared with that pertaining to the zero-order field (let us say $P_{AP0}$).

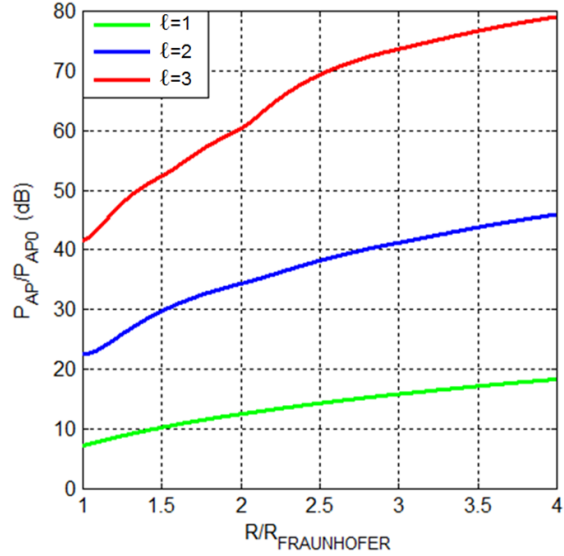

Fig. 3. Aperture power required to generate different OAM states as a function of the link distance (for $a_1=a_2=5\lambda$).

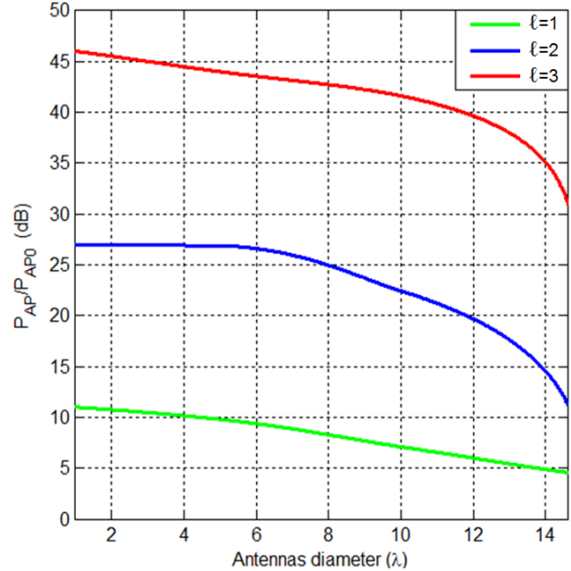

Fig. 4. Aperture power required to generate different OAM states as a function of the antenna size (for $a_1=a_2$ and $R=R_F$).

Two kinds of analysis have been carried out as discussed in the following.

As a first set of numerical trials, we fixed $a_1=a_2=5\lambda$, and repeatedly solved problem (20) for $R\epsilon[R_F, 4R_F]$. The achieved results are summarized in Fig. 3, wherein the ratio between the aperture powers $P_{AP}$ and $P_{AP}0$ is plotted as a function of the ratio between the actual and Fraunhofer link distances. As it can be seen, two important statements can be made:

- for a fixed link distance, a *small* increase of $|\ell|$ corresponds to a *huge* increase of the required aperture power;

- whatever the $\ell$ value, the energy required to generate a given OAM state is a monotonically increasing function of the link distance. For $R\approx 4R_F$, even if $\ell$ is low, the curves already achieve extremely large values, e.g., the power required to generate an angular variation of order $\ell=3$ is about 80-dB higher than the one needed in case $\ell=0$.

These results confirm that it is preferable avoiding adoption of OAM antennas in long-range communication links and that, anyway, it is not possible to realize arbitrarily-large values of $\ell$ (and hence "arbitrarily-large" channel capacities). Notably, the approach allows a quantitative evaluation of the price to be paid. All these considerations agree with the ones made in [7],[8] as well as with the outcomes of the few experimental proofs provided up to now. Moreover, it is worth noting that the results in Fig. 3 (green curve) seem to be coherent with the ones in [20]. In the latter paper, in fact, authors simultaneously employ an amplifier on the $\ell=\pm1$ antennas and an attenuator on the $\ell=0$ antenna (for a total of 24 dB) in order to equalize received power.

As a second set of numerical trials, by using $a_1=a_2$ and $R=R_F$, we repeatedly solved problem (20) for $2a_1 \in (1\lambda, 15\lambda)$. The achieved results are summarized in Fig. 4, wherein the ratio between the aperture powers $P_{AP}$ and $P_{AP0}$ is plotted as a function of the antennas size. The synthesis outcomes reveal two important circumstances:

- whatever the antennas size, a *huge* increase of the aperture power is required to realize a *small* increase of $|\ell|$;

- whatever the $\ell$ value, the input power is a monotonically decreasing function of the antenna size. This is due to the fact that increasing $a_2$ ($a_1$) allows dealing with larger $\theta_1$ values, thus mitigating the effects of the 'holes' present in the field pattern around the boresight direction.

All the above results, based on a rigorous quantitative setting, confirm the OAM antennas' limitations identified in the previous paragraphs and Subsections and, again, agree with results in [7],[8],[12].

## VI. CONCLUSIONS

The classical Aperture Antennas Theory has been exploited to give a further point of view on the actual capabilities of OAM antennas. Such an approach suggests the use of Hankel transforms as a convenient and simple tool for understanding the ultimate limitations of these radiating systems. The adopted theoretical framework, equipped with the SVD analysis of the 'radiation operators' at hand, allows stating that the adoption of OAM antennas is not convenient in very-long-range communication links. Moreover, the proposed tools allow identifying the OAM systems' number of *degrees of freedom* and to quantitatively understand how performances of these antennas vary with order, varying sizes and distances.

The overall outcomes indicate that for a fixed size of the source a huge price is paid for generating vortices corresponding to higher and higher values of $|\ell|$, and that an 'arbitrarily large' multiplication of channels in the far-field zone is de facto unfeasible.

The achieved results do not go against the possibility of fruitfully adopting OAM antennas' in near field links. In fact, by using an extended receiving antenna in the near-field zone it could be possible to increase the channel capacity with respect to the case where the same antenna is in the far-field region. On the other side, for a really meaningful assessment, the realized number of channels should be then compared with alternative MIMO/multibeam possibilities. In fact, the receiving and/or transmitting areas could be partitioned in smaller antennas.

In such an interesting study (implicitly suggested by one Reviewer) the above tools are still of interest. In fact, according to the Aperture Antenna Theory, the evaluation of the field at any distance from the aperture can be carried out by just multiplying the aperture field's spectrum by a proper exponential function, and performing an inverse transform [21].

## APPENDIX I

The aim of this Appendix is to derive the expressions of the singular functions and the singular values of the operator $A_\ell$.

In order to diagonalize $A_\ell$ in a simple and efficient fashion, it proves convenient (see [22],[23]) to introduce the auxiliary operator $\bar{A}_\ell$ defined by:

$$\bar{A}_\ell f_\lambda = \int_0^1 f_\lambda(\rho) J_\lambda(\beta ak\rho)\sqrt{\beta ak\rho}\, d\rho \qquad (21.a)$$

with:

$$\bar{A}_\lambda : f_\lambda \in L_2(0,1) \to \bar{A}_\lambda f_\lambda \in L_2(0,1) \qquad (21.b)$$

Notably, this operator is equal to the one extensively studied in [22] through an eigenvalue decomposition (wherein the eigenfunctions represent the GSPW functions).

It is also interesting to note two crucial properties of $\bar{A}_\ell$:

- as it is a compact self-adjoint operator, its eigenvalues are real [37];

- as it is a normal operator, its singular values are equal to the amplitude of its eigenvalues [38].

Therefore, the singular values of $\bar{A}_\ell$ results equal (but for a change of sign for negative values of the eigenvalues) to its eigenvalues, which have been analytically derived in [22]. On the other side, it is indeed very difficult to extract from [22] the results of interest herein, so that it makes sense to provide an alternative easier approach.

In this respect, it is useful to note that, for any fixed size of the source, one can compute the singular values and singular functions of $A_\ell$ through numerical discretization by first computing the SVD of $\overline{A}_\ell$, i.e., the functions and scalars $\{\overline{v}_{\ell,n}, \overline{\sigma}_{\ell,n}, \overline{u}_{\ell,n}\}$ such that:

$$\overline{A}_\lambda \overline{v}_{\lambda,n} = \overline{\sigma}_{\lambda,n} \overline{u}_{\lambda,n}$$
$$\overline{A}_\lambda^+ \overline{u}_{\lambda,n} = \overline{\sigma}_{\lambda,n} \overline{v}_{\lambda,n}$$

(22)

and then using:

$$u_{\lambda,n} = \frac{\overline{u}_{\lambda,n}}{\sqrt{k}} \qquad (23.a)$$

$$v_{\lambda,n} = \frac{\overline{v}_{\lambda,n}}{\sqrt{\rho}} \qquad (23.b)$$

$$\sigma_{\lambda,n} = \frac{\overline{\sigma}_{\lambda,n}}{\sqrt{\beta a}} \qquad (23.c)$$

In fact, it is:

$$A_\lambda v_{\lambda,n} = \int_0^1 J_\lambda(\beta ak\rho) \frac{\overline{v}_{\lambda,n}}{\sqrt{\rho}} \rho d\rho =$$

$$= \frac{1}{\sqrt{\beta ak}} \int_0^1 J_\lambda(\beta ak\rho)\sqrt{\beta ak\rho} \, \overline{v}_{\lambda,n} d\rho = \frac{1}{\sqrt{\beta ak}} \overline{A}_\lambda \overline{v}_{\lambda,n} =$$

(24)

$$= \frac{1}{\sqrt{\beta ak}} \overline{\sigma}_{\lambda,n} \overline{u}_{\lambda,n} = \sigma_{\lambda,n} \frac{\overline{u}_{\lambda,n}}{\sqrt{k}} = \sigma_{\lambda,n} u_{\lambda,n}$$

Summarizing, in order to understand the typical behavior of singular values and singular functions of the different $A_\ell$ operators, one needs:

i. the SVD of the auxiliary $\overline{A}_\lambda$ operators;

ii. some robust and accurate implementation of (23.a), (23.b).

Obviously, step (i) does not require any particular care (but for a sufficiently large number of discretization points). Step (ii) requires instead some trick in order to avoid the difficulties related to the presence of singularities at the denominator. Such a difficulty can be however circumvented by using the simple method described in the Appendix II.

## APPENDIX II

The aim of this Appendix is to describe a simple method to avoid the difficulties related to the presence of singularities at the denominator of (23.a) and (23.b).

The divisions required by (23.a) and (23.b) can be numerically performed in an accurate and reliable fashion by means of the following three-steps procedure:

1. multiply the functions $\overline{u}$ and $\overline{v}$ by $k^{1/2}$ and $\rho^{1/2}$, respectively;

2. perform a polynomial fitting of the functions coming out from step 1;

3. erase the constant term of the polynomial coming out from step 2, and lower by one the order of all other terms.

In fact, since the functions coming out from step 1 have a zero in the origin, they cannot contain a constant term and hence step 3 is equivalent to respectively divide them by $k$ and $\rho$.

## REFERENCES

[1]   A. E. Willner, "Communication with a twist," *IEEE Spectrum*, vol. 53, n. 8, pp. 34-39, 2016.

[2]   F. Tamburini, E. Mari, A. Sponselli, B. Thidé, A. Bianchini, and F. Romanato, "Encoding many channels on the same frequency through radio vorticity: First experimental test," *New Journal of Physics*, vol. 14, 033001, 17 pages, 2012.

[3]   Y. Yan, G. Xie, M. P. J. Lavery, H. Huang, N. Ahmed, C. Bao, Y. Ren, Y. Cao, L. Li, Z. Zhao, A. F. Molisch, M. Tur, M. J. Padgett, and A. E. Willner, "High-capacity millimetre-wave communications with orbital angular momentum multiplexing," *Nature Communications*, vol. 5, p. 4876, 2014.

[4]   N. Bozinovic, Y. Yue, Y. Ren, M. Tur, P. Kristensen, H. Huang, A. E. Willner, and S. Ramachandran, "Terabit-scale orbital angular momentum mode division multiplexing in fibers," *Science*, vol. 340, n. 6140, pp. 1545-1548, 2013.

[5] M. Pu, X. Li, X. Ma, Y. Wang, Z. Zhao, C. Wang, C. Hu, P. Gao, C. Huang, H. Ren, X. Li, F. Qin, J. Yang, M. Gu, M. Hong, and X. Luo1, "Catenary optics for achromatic generation of perfect optical angular momentum," *Science Advances*, vol. 1, n. 9, e1500396, 2015.

[6] X. Cai, J. Wang, M. J. Strain, B. Johnson-Morris, J. Zhu, M. Sorel, J. L. O'Brien, M. G. Thompson, S. Yu, "Integrated Compact Optical Vortex Beam Emitters," *Science*, vol. 338, n. 6105, pp. 363-366, 2012.

[7] O. Edfors and A. J. Johansson, "Is orbital angular momentum (OAM) based radio communication an unexploited area?," *IEEE Transactions on Antennas and Propagation*, vol. 60, no. 2, pp. 1126–1131, 2012.

[8] M. Tamagnone, C. Craeye, and J. Perruisseau-Carrier, "Comment on 'Encoding many channels on the same frequency through radio vorticity: First experimental test'," *New Journal of Physics*, vol. 14, p. 118001, 2012.

[9] F. Tamburini, B. Thidé, E. Mari, A. Sponselli, A. Bianchini, and F. Romanato, "Reply to Comment on 'Encoding many channels on the same frequency through radio vorticity: First experimental test'," *New Journal of Physics*, vol. 14, p. 118002, 2012

[10] M. Tamagnone, C. Craeye, and J. Perruisseau-Carrier, "Comment on Reply to Comments on 'Encoding many channels on the same frequency through radio vorticity: First experimental test'," *New Journal of Physics*, vol. 15, p. 118001, 2013.

[11] F. Spinello, G. Parisi, F. Tamburini, G. Massaro, C. G. Someda, M. Oldoni, R. A. Ravanelli, F. Romanato, and E. Mari, "High-order vortex beams generation in the radio-frequency domain," *IEEE Antennas and Wireless Propagation Letters*, vol. 15, pp. 889-892, 2016.

[12] C. Craeye, "On the transmittance between OAM antennas," *IEEE Transactions on Antennas and Propagation*, vol. 64, no. 1, pp. 336-339, 2016.

[13] M. Barbuto, F. Trotta, F. Bilotti, and A. Toscano, "Circular polarized patch antenna generating orbital angular momentum," *Progress In Electromagnetics Research*, vol. 148, pp. 23-30, 2014.

[14] A. Cagliero, A. De Vita, R. Gaffoglio, and B. Sacco, "A new approach to the link budget concept for an OAM Communication Link," *IEEE Antennas and Wireless Propagation Letters*, vol. 15, pp. 568-571, 2016.

[15] L. Cheng, W. Hong, and Z. Hao, "Generation of electromagnetic waves with arbitrary orbital angular momentum modes," *Scientific Reports*, vol. 4, article n. 4814, 2014.

[16] D. K. Nguyen, O. Pascal, J. Sokoloff, A. Chabory, B. Palacin, N. Capet, "Antenna gain and link budget for waves carrying orbital angular momentum," *Radio Science*, vol. 50, n. 11, pp. 1165–1175, 2015.

[17] A. Hellemans, "A new twist on radio waves," *IEEE Spectrum*, vol. 49, n. 5, pp. 16- 18, 2012.

[18] R. Gaffoglio, A. Cagliero, A. De Vita, and B. Sacco, "OAM multiple transmission using uniform circular arrays: numerical modelling and experimental verification with two digital television signals," in print on *Radio Science*, 2016.

[19] M. Oldoni, F. Spinello, E. Mari, G. Parisi, C. G. Someda, F. Tamburini, F. Romanato, R. A. Ravanelli, P. Coassini, and B. Thidé, "Space-division demultiplexing in Orbital-Angular-Momentum-based MIMO radio systems," *IEEE Transactions on Antennas and Propagation*, vol. 63, n. 10, pp. 4582-4587, 2015.

[20] F. Tamburini, E. Mari, G. Parisi, F. Spinello, M. Oldoni, R. A. Ravanelli, P. Coassini, C. G. Someda, B. Thidé, and F. Romanato, "Tripling the capacity of a point-to-point radio link by using electromagnetic vortices," *Radio Science*, vol. 50, pp. 501–508, 2015.

[21] C. A. Balanis, *Antenna Theory, Analysis, and Design*, §12.9 "Fourier Transforms in Aperture Antenna Theory," pp. 701-707, John Wiley & Sons, 2005.

[22] D. Slepian, "Prolate spheroidal wave functions, Fourier analysis and uncertainty — IV: Extensions to many dimensions; generalized prolate spheroidal functions," *The Bell System Technical Journal*, vol. 43, n. 6, pp. 3009-3057, 1964.

[23] G. T. di Francia, "Degrees of freedom of an image," *Journal of the Optical Society of America*, vol. 59, n. 7, pp. 799-804, 1969.

[24] R. Bracewell, "The Hankel Transform", *The Fourier Transform and Its Applications*, pp. 335-340, New York: McGraw-Hill, 2000.

[25] R. Piessens, "The Hankel Transform," *The Transforms and Applications Handbook*, §9, Alexander D. Poularikas, Boca Raton: CRC Press LLC, 2000.

[26] O. M. Bucci, T. Isernia, and A. F. Morabito, "An effective deterministic procedure for the synthesis of shaped beams by means of uniform-amplitude linear sparse arrays," *IEEE Transactions on Antennas and Propagation*, vol. 61, n. 1, pp. 169-175, 2013.

[27] M. Abramowitz and I. A. Stegun, "Bessel Functions *J* and *Y*," *Handbook of Mathematical Functions with Formulas, Graphs, and Mathematical Tables*, §9.1, pp. 358-364, New York: Dover 1972.

[28] R. L Phillips and L. C. Andrews, "Spot size and divergence for Laguerre-Gaussian beams of any order," *Applied Optics*, vol. 22, n. 5, pp.643-644, 1983.

[29] M. Bertero and P. Boccacci, "Singular Value Decomposition (SVD)", *Introduction to Inverse Problems in Imaging*, §9, pp. 220-246, CRC Press, 1998.

[30] O. M. Bucci, T. Isernia, and A. F. Morabito, "Optimal synthesis of circularly symmetric shaped beams," *IEEE Transactions on Antennas and Propagation*, vol. 62, n. 4, pp. 1954-1964, 2014.

[31] D. R. Rhodes, "On the quality factor of strip and line source antennas and its relationship to superdirectivity ratio," *IEEE Transactions on Antennas and Propagation*, vol. 20, n. 3, pp. 318-325, 1972.

[32] O. M. Bucci, T. Isernia, and A. F. Morabito, "Optimal synthesis of directivity constrained pencil beams by means of circularly symmetric aperture fields," *IEEE Antennas and Wireless Propagation Letters*, vol. 8, pp. 1386-1389, 2009.

[33] E. Hille and J. D. Tamarkin, "On the characteristic values of linear integral equations," *Acta Math.*, vol. 57, pp. 1-76, 1931.

[34] Bucci, O. M., and T. Isernia. "Electromagnetic inverse scattering: Retrievable information and measurement strategies." *Radio Science,* vol. 32, no. 6 pp. 2123-2137, 1997.

[35] O. M. Bucci and G. Franceschetti, "On the degrees of freedom of scattered fields," *IEEE Transactions on Antennas and Propagation*, vol. 37, n. 7, pp. 918-926, 1989.

[36] O. M. Bucci, C. Gennarelli, and C. Savarese, "Representation of electromagnetic fields over arbitrary surfaces by a finite and nonredundant number of samples," *IEEE Transaction on Antennas and Propagation*, vol. 46, n. 3, pp. 351-359, 1998.

[37] J. B. Conway, "The diagonalization of compact self-adjoint operators," *A Course in Functional Analysis*, §5, pp. 46-49, Springer, 1990.

[38] A. Ruhe, "On the closeness of eigenvalues and singular values for almost normal matrices," *Linear Algebra and its Applications*, vol. 11, n. 1, pp. 87-93, 1975.

"